\documentclass[a4paper,11pt]{article}
\usepackage{fullpage}
\usepackage{times}
\usepackage{subfigure}
\usepackage{vmargin}
\usepackage{fancyheadings}
\usepackage[cp1251]{inputenc}
\usepackage[english]{babel}
\usepackage{graphicx}
\usepackage{color}
\usepackage{cite}
\usepackage{amsmath,amssymb,amsthm,enumerate,bbm,color,verbatim,setspace}

\newtheorem{rem}{Remark}[section]

\newtheorem{St}{Statement}[section]
\newtheorem{example}{Example}[section]

\date{}
\title{Optimal nonlinear filtering of quantum state}
\author{V.I. Man'ko$^{1,2}$ and L.A. Markovich$^{2,3,4*}$\\
      $^1$P. N. Lebedev Physical Institute, Russian Academy of Sciences\\
Leninskii Prospect 53, Moscow 119991, Russia\\
$^2$Moscow Institute of Physics and Technology\\
 Institutskii Per. 9, Dolgoprudny Moscow Region 141700, Russia\\
$^3$Institute for information transmission problems, Moscow,\\ Bolshoy Karetny per. 19, build. 1, Moscow 127051, Russia\\
$^4$V. A. Trapeznikov Institute of Control Sciences, Moscow,\\Profsoyuznaya 65, 117997 Moscow, Russia\\
$^*$Corresponding author e-mail: kimo1@mail.ru}
\date{}
\begin{document}
\maketitle
\pagenumbering{arabic}
\begin{abstract}\noindent
 The aim of this work is to give an interpretation of the linear or nonlinear
filtration problem as a quantization problem. Based on this, we extend the optimal filtering equation known from the Stratonovich filtering theory
 on the quantum process case. In contrast to the Kalman filter often used in such problems, the Stratonovich equation provides the optimal solution for nonlinear observation models. The observation model is based on an indirect measurement method, where the measurement is performed on an ancilla
system that is interacted with an unknown one. The latter model is interpreted in case of a single - qudit system. Since the measured random variables and experiment are fundamentally different from each other, the obtained single - qudit observation model has a substantially different physical nature from the two-qubit one. However, in both problems, the   Stratonovich optimal filter equation is equally applicable.
\end{abstract}
\medskip


\section{\label{sec:level1}Introduction}
\par With the growth of experimental base of quantum mechanics  there is an increasing awareness of extension of classical probability theory, filtration and control results to the quantum problems. Measuring physical quantities, experimentalists face with such difficulties as the optimal measurement problem (see \cite{Edwards2005OptimalQF}) or the quantum filtering problem (see \cite{Belavkin2}). For example, the observable random variable may contain a mixture of the useful signal with noise. Also, the system of interest  may not be available for the observation and the information about it is obtained through an observable system which is somehow coupled with the original one. In both cases, the knowledge about the system of interest must be "filtered" from the measurements done on the observable one.
\par There has been recent interest in quantum filtering and state estimation problems in quantum information theory and quantum
control (cf. \cite{Armen,Barchielli,Bouten}). Last experimental and theoretical advances in quantum technology provide a
strong motivation for the quantum control  including the engineering of quantum states, the stability theory, the quantum
error correction, the robust control and quantum networks (cf. \cite{Ahn,Sarovar,Belavkin1,Doherty,James}). The quantum control  plays a
fundamental role in the development of new quantum technologies like the quantum computation. For example, the quantum filtering in coherent states is introduced in \cite{Gough}
and the tomographical approach is used for the signal recognition and denoising in a number of recent papers (cf. \cite{Aguirre,Mendes,Mendes2}).
\par For quantum state estimation, is known, one has to give the measurement strategy that is used to get information, and the
estimator mapping the measurement data to the state space.
In this paper, we consider the weak measurement model, where unobservable quantum
system is coupled with an ancilla system  (a probe system) that can be measured.
A von Neumann measurement is used, described by  a set of projection operators $\{P_n = |n\rangle\langle n|\}$. Each operator describes what happens on one of  possible outcomes
of the measurement. The method is based on a set of copies of the same system being in the same state.
The measurement scheme described above is based on two-qubit systems. This is due to a fast development of quantum technologies such as quantum computers,
 quantum cryptography and teleportation. These technologies promise in the future to lead to a revolution in computation and communication.
\par The quantum computing is built by elementary processing elements, namely, quantum bits (qubits). It is known that classical computers elements (bits) take only two values, logic zero and logic one. On the contrary, the qubits as quantum objects can be located also in a coherent superposition of these two states. Thus, they describe the intermediate state, between the logic zero  and one. By measuring the qubit, we get zero or one with some probabilities.
The quantum computers will be able in a finite time to solve problems, to the solution of which the classical supercomputers will take much more time.
An implementation of such computer has several obstacles. Quantum states of ions, electrons and Josephson junctions used as qubits are extremely unstable and cannot be kept in the same state
for a long time. For the implementation of computational algorithms one needs  a set of interacted qubits in some particular state. In recent years,
the stability of the states of qubits has been increased, but its improvement is still  extremely complicated problem, especially for the multipartite
systems with a large number of qubits.
\par One way to solve this problem is to use single - qudit systems as the quantum objects  instead of qubits (cf. \cite{Kiktenko,Kiktenko2}). Such single - qudit systems have more states than the two-qubit ones.
 Thus, the number of particles (or subsystems) in the system is greatly reduced. A wide number of papers exist  devoted to the study of various kinds of characteristics of quantum correlations in systems
 with subsystems such as the two-qubit systems. Such quantum system can have a correlation responsible for the entanglement phenomenon (cf. \cite{schredinger:35}) and for the violation of Bell's inequality (cf. \cite{Horn}). These correlations may also be responsible for the quantum discord (cf. \cite{Yurkevich,Mscord}). However, in literature the systems without  subsystems (i.e. one qutrit or qudit)  are payed less attention. Recently, in \cite{Chernega:2008,Chernega:14,ManMan,Manko1} it was shown that the quantum properties of systems without subsystems can be formulated using the method of an invertible  mapping. In \cite{Manko1} the entanglement  concept and correlations in the single - qudit state are discussed.
  Using the latter mapping, the notion of the separability and the entanglement is extended in \cite{Markovich3} to the case of the single - qudit $X$-state with $j=3/2$.
\par Thus, the state space model for a particular quantum system based on the weak measurements can be extended to the case of a single - qudit system, where the "artificial" unobservable qubit and ancilla qubit are introduced. The proposed observation models both for the qubits and for the one qudit systems are often nonlinear. Therefore, the widely known Kalman filtering method \cite{Kalman} is not applicable in this case. Hence, a method is required that performs good both in linear and nonlinear cases.
\par The problems related to the quantum state estimation and the quantum state filtering are fundamental for the quantum information
theory and quantum control. In \cite{Ruppert} the well-known procedure in the field of classical
control theory, namely, the Kalman filtration (see \cite{Kalman}) has been applied to the quantum filtering area. The problem of the filtering of unknown signals from the mixture with noise is well studied in classical probability theory. The Kalman filter provides the optimal solution for the linear recursive model of the observation.
However, for nonlinear models that appear often in  practice the Kalman filter is not applicable.
\par As it was shown in \cite{Mancini,Asorey} the quantum states can be described by the fair probability distribution function called the quantum tomogram. It means that  density matrices or vectors in the Hilbert spaces can be mapped onto probability distributions. In view of this  filtration procedures known in classical probability theory can be applicable to the quantum problem.
\subsection{Contributions of this paper}
\par The paper is aimed as a conceptually simple and intuitive approach to the Stratonovich's optimal filtration procedure. We
deal with both the linear and nonlinear filtering, but final results concern the nonlinear case. First, we state the problem of filtering of a partially observable Markov quantum measurement process, where the optimal estimator of an unobserved component is based on observations. To this end, the general filtering equation introduced in \cite{Stratonovich:60,Dobrovidov:1983,Dobrovidov:2012} is used. In the author's paper \cite{MarLit} it is proved that the general filtering equation that gives the optimal Bayesian estimator is the Kalman filter in the case of linear model. Note that the general filtering equation does not contain  explicit probabilistic characteristics of the unknown unobservable sequence. We find the optimal state estimate by only observable quantities.
Our second aim  is to propose a quantum measurement model  for the single - qudit state. The optimal filtering method proposed for the two-qubit models can be extended to the latter case.  The construction of such type of observation models is useful in the light of possible practical use of the single - qudit systems.
For example, one can measure the population on only  certain  levels of a multi-level atom, while other levels are not available to measure.
This may be due to their short lifetime or a diversity in the frequency band reception. Then the measurement scheme described below allows to extract information about inaccessible levels  by measuring only available ones. %Finally, we introduce the tomographic representation of the observation model and show its change at each time step of the measurement. The time evolution of the quantum information is shown and it illustrates the  correlation changes in the quantum system.
\par The paper is organized as follows. In Sec.~\ref{sec:12_0}  we give a brief overview of results known for the classical nonlinear filtering problem. The Stratonovich's optimal approach for nonlinear processes is discussed  in details. In Sec.~\ref{sec:12_1} we recall the notion of the weak measurement for the system of two qubits.
We state the problem of filtering of the partially observable two-qubit system and the nonlinear measurement scheme is constructed. The Stratonovich's approach used for the classical filtering problem is interpreted as a quantization problem. It is shown that by means of the optimal filtering equation one can find the optimal solution of the nonlinear quantum filtering problem without  linearization procedures, see \cite{Ruppert} among others. In the next section the method of the invertible mapping is used to obtain the non-linear measurement model for the single - qudit system. The physical meaning of correlations in such system and the construction of possible experiment are given.
%In Sec.~\ref{sec:12_4} the observation model is rewritten in term of the tomogram. The latter approach is used to write the Shannon entropies and information depending on a time step of the observation model. Thus, it is possible to observe the time evolution of the  information and hence, the correlation change in the quantum system. Section \ref{sec:12_5} concludes.
\section{Filtration problem}\label{sec:12_0}
Here, we briefly present the results known for the classical nonlinear filtering problem (cf. \cite{Stratonovich:60,Dobrovidov:1983,Dobrovidov:2012}).
Suppose that a partially observable  Markov random process $(s_k,x_k)_{k\geq1}$, where the sequence $s = (s_k)_{k\geq1}$  is unobservable and the sequence $x=(x_k)_{k\geq1}$ is observable,
is statistically coupled by the conditional density $f(x_k|s_k)$. The conditional density of observations depends on the observation model and  the distribution of the observation noises
$\eta_k$. Hence, depending on the form of the observation model and the noise distribution one has different classes of filtering problems. In this paper, we consider the exponential family
of the conditional densities that are commonly used in practice, i.e.
\begin{eqnarray}\label{exponetfamily}f(x_k|s_k)=\widetilde{C}(s_k)h(x_k)\exp(T(x_k)Q(s_k)),
\end{eqnarray}
where $\widetilde{C}(s_n)$ is a normalization constant and $h(x_k),T(x_k),Q(s_k)$ are known functions.
\par We proceed to construct the filtering algorithm for an unobservable random sequence $s = (s_k)_{k\geq1}$, based on observations of the sequence $x=(x_k)_{k\geq1}$
that are connected by the following nonlinear  expression
\begin{eqnarray}\label{12_33}
x_k&=&\varphi(s_k,\eta_k),
\end{eqnarray}
where $(\eta_k\in\mathbb{R})_{k\geq1}$ is an independent identically distributed random sequence, $(s_k)_{k\geq1}$ is a Markov sequence and $\varphi$ is some reversible function. Realizations of the random variables $s_k\in\mathcal{S}_k\subseteq\mathbb{R}$ and $x_k\in\mathcal{X}_k\subseteq\mathbb{R}$ are denoted by $s_1^k=(s_1,\ldots,s_k)^T$ and $x_1^k=(x_1,\ldots,x_k)^T$, respectively. The assumption \eqref{exponetfamily} imposes some restrictions on $\varphi$ and $\eta_k$.
There must be a unique solution of  \eqref{12_33}, i.e.
\begin{eqnarray}\label{12_35}
\eta_k&=&\phi(s_k,x_k),
\end{eqnarray}
and  the probability density of noise $p_{\eta_k}(y_k)$
must be such, that after the substitution, the conditional density $f(x_k|s_k)$ were from the exponential class.
%\par Let us define the random sequence  $\vartheta_k=Q(s_k)$, where the random variable  $s_k$ is related to $\vartheta_k\in\Theta_k\subseteq\mathbb{R}$ by some one-to-one function $Q:\mathcal{S}_k\rightarrow\Theta_k$. The random sequence $(\vartheta_k)_{k\geq1}$ is also a Markov sequence. We choose the quadratic loss function
%\begin{eqnarray}L(s_k,\hat{s}_k)=\left(Q(s_k)-Q(\widehat{s}_k)\right)^{T}\left(Q(s_k)-Q(\widehat{s}_k)\right)^{T}\end{eqnarray}
%and the optimal Bayesian estimator that  minimizes the risk function
%\begin{eqnarray}R(\hat{\vartheta}_k)=E\left(\left(Q(s_k)-Q(\widehat{s}_k)\right)^{T}\left(Q(s_k)-Q(\widehat{s}_k)\right)^{T}\right)\label{12_01}\end{eqnarray}
%is equal to the mathematical expectation $\hat{\vartheta}_k=E\left(\vartheta_k|x_1^k\right)$.
\par In case when \eqref{12_33} has the recursive linear form
\begin{eqnarray}\label{12_0}s_k&=&as_{k-1}+b\xi_k,\\
x_k&=&As_k+B\eta_k,\nonumber
\end{eqnarray}
where $s_k,x_k\in\mathbb{R}$ for all $k$, $\xi_k$ and $\eta_k$ are mutually independent random variables with the standard Gaussian distribution,
\begin{eqnarray*}   s_0\in\mathcal{N}(0,\widetilde{\sigma}^2),\quad \widetilde{\sigma}^2=b^2/(1-a^2),\quad
s_k\in \mathcal{N}(0,1),\quad  k=1,2,3\ldots,
\end{eqnarray*}
coefficients $A,B,a,b$ are given by real numbers and $|a|<1$, then the Kalman filter is applied as an optimal filtration method \cite{Kalman}.
The Kalman estimator is optimal with regard to the minimum of the risk function
\begin{eqnarray}R(\hat{s}_k)=E\left(\left(s_k-\widehat{s}_k\right)\left(s_k-\widehat{s}_k\right)^{T}\right)\label{12_01}\end{eqnarray}
and is equal to the mathematical conditional  expectation $\hat{s}_k=E\left(s_k|x_1^k\right)$.
However, for all the other cases, the Kalman filter doesn't provide an optimal solution to the filtration problem and cannot be applied.
\par The optimal approach for nonlinear processes is proposed in \cite{Stratonovich:60}.
To estimate $s_k$ the optimal Bayesian estimator in the form of the conditional mean
\begin{eqnarray}\label{theta}\widehat{s}_k=\mathsf{E}(s_k|x_1^k)=\int_{\mathcal{S}_k}s_kw_k(s_k|x_1^k)ds_k,
\end{eqnarray}
has been used. The  $w_k(s_k|x_1^k)$ is the posterior probability density function  that satisfies the Stratonovich's recurrence equation (cf. \cite{Strat66,Kushner}) given by
\begin{eqnarray*}&& w_{1}(s_{1}|x_1)=\frac{f(x_1|s_1)p(s_1)}{\int _{\mathcal{S}_1}f(x_1|s_1)p(s_1)ds_1},\\\nonumber
&&w_k(s_k|x_1^k)=\frac{f(x_k|s_k)}{f(x_k|x_1^{k-1})}\int_{\mathcal{S}_{k-1}}p(s_k|s_{k-1})w_{k-1}(s_{k-1}|x_1^{k-1})ds_{k-1},\!\!\!\quad n\geq2.\label{555}
\end{eqnarray*}
Here, we denote  the transition probability density function of the Markov sequence $(S_k)_{k\geq1}$ as $p(s_k|s_{k-1})$ and $f(x_k|x_1^{k-1})$, $f(x_k|s_k)$ denote conditional densities.
\par Since the posterior density $w_k(s_k|x_1^k)$ depends on the
unknown prior distribution function $p(s_1)$ and the transition probability $p(s_k|s_{k-1})$ of the
Markov sequence $(s_k)_{k>1}$, we cannot use formula the latter formula to estimate $\widehat{s}_k$.
\par To overcome this problem the optimal filtering equation is proposed in \cite{Dobrovidov:1983,Dobrovidov:2012}. The equation is derived under the assumption that the conditional density $f(x_k|s_k)$ belongs to the exponential family of distributions.
\begin{rem}Note that this assumption is not necessary. One can derive an analogical filtering equation for other classes of densities.\end{rem}
The optimal filtration equation is (cf. \cite{Dobrovidov:2012})
\begin{eqnarray}\label{D}\mathsf{E}(s_k|x_1^k)\cdot T'_{x_k}(x_k)&=&
\left(\ln\left(f(x_k|x_1^{k-1})/h(x_k)\right)\right)'_{x_k}.
\end{eqnarray}
It is easy to see that equation \eqref{D} does not contain explicit probabilistic characteristics $p(s_1)$ and $p(s_k|s_{k-1})$ of the unknown sequence  $(s_k)$. This allows us to find the optimal estimator \eqref{theta} (that minimizes the mean squared deviation from the true value of an unobservable $s_1^k$) knowing only observable quantities of $x_1^k$.
However, the latter equation contains the logarithmic derivative of the unknown conditional density $f(x_k|x_1^{k-1})$ which characterizes the signal. The logarithmic probability density function  (pdf) derivative  is the ratio of the derivative of the pdf to the pdf itself. To estimate them the gamma product kernel estimators by multivariate dependent data are introduced by Markovich in \cite{Mar_2} (see example 4.3 on p.14).
\par One can take the Gaussian density
\begin{eqnarray}\label{f}f(x_k|s_k)=\frac{1}{\sqrt{2\pi}B}\exp\left(-\frac{(x_k-As_k)^2}{2B^2}\right)
\end{eqnarray}as an example of exponential family \eqref{exponetfamily}.
Then the observation model is defined by the linear equation
\begin{eqnarray}\label{0_0}
x_k&=&As_k+B\eta_k,
\end{eqnarray}
where $\{\eta_k\}$ are independent identically distributed random variables with the Gaussian distribution and coefficients $A$ and $B$ are real numbers.
Hence,   equation  \eqref{D} can be rewritten in the special form
\begin{eqnarray}\label{th}\mathsf{E}(s_k|x_1^k)&=&\frac{B^2}{A}\frac{f'_{x_k}(x_k|x_1^{k-1})}{f(x_k|x_1^{k-1})}+\frac{x_k}{A}
\end{eqnarray}
which is the Kalman filter.
\begin{rem}The exact coincidences of the triple: the optimal filtering
equation \eqref{D} of the unobservable Markov sequence $(s_k)$ defined by a linear equation with a Gaussian noise, the Kalman filter and the conditional expectation $\mathsf{E}(s_k|x_1^k)$ defined by Theorem of normal correlation \cite{ShiryaevLiptser:2001} is proved by the author in \cite{MarLit}.
Thus, the optimal  filtering equation is nothing else but the Kalman filter in case of linear model \eqref{12_0}. However,
the general  filtering equation provides the optimal solution for nonlinear processes in contrast to the Kalman filter that cannot be applied to nonlinear models.\end{rem}
\section{The weak measurement}\label{sec:12_1}
\par Obviously, the approach introduced above can be applied to arbitrary Markovian pair, regardless of the nature of the measurements. Therefore, we want to apply the latter theory  to quantum filtration. In this article, we consider the discrete time case of the indirect measurement (see \cite{Ruppert}).
We suppose that the unobservable and the observable measurements of quantum systems are  quantum bits.
Under the weak measurement it means that the projective measurements are done on the extra ancilla system (the probe or the measurement device) that is in state $\theta_M(k)=[\theta_{M_1}(k),\theta_{M_2}(k),\theta_{M_3}(k)]^T$ coupled with the system $\theta_S(k)=[\theta_{S_1}(k),\theta_{S_2}(k),\theta_{S_3}(k)]^T$ that we are interested in. Two Bloch's vector representations of the latter states are
\begin{eqnarray}\label{11}&&\rho_M(k)=(I+\theta_{M}(k)\sigma^M)/2,\quad \rho_S(k)=(I+\theta_{S}(k)\sigma^S)/2,
\end{eqnarray}
where the $\sigma^S$ and $\sigma^M$ are symbolic vectors constructed from Pauli operators acting on Hilbert spaces $H_S$ and $H_M$, respectively.
The indirect measurement is proceeded  by the following way. At the time step $k$ we prepare the ancilla qubit in a known state and couple it to an unknown system.
The composite system is represented by the 4-dimensional square density matrix $\rho_{S+M}(k)$. Let us take it as a direct product of two latter states, i.e. $\rho_{S+M}(k)=\rho_{S}(k)\otimes\rho_{M}(k)$.
Both qubits evolve at sampling time $h$ according to  bipartite dynamics. At the end we do the von Neumann measurement on the ancilla qubit.
Generally, the von Neumann measurement is the measurement of Pauli operators. The algorithm is repeated at the next time step $k+1$ .
\par For example, if we are interested in the measurement of the observable $\sigma_x$, then possible
outcomes are its eigenvalues $(\pm1)$. The probabilities of two different outcomes
\begin{eqnarray*}A_x&=&I\otimes\sigma_x=\left(1/2\right)|+\rangle\langle+|+\left(-1/2\right)|-\rangle\langle-|,\\
A_x|+\rangle\langle+|&=&\left(1/2\right)|+\rangle\langle+|,\quad A_x|-\rangle\langle-|=\left(-1/2\right)|-\rangle\langle-|
\end{eqnarray*}
of the von Neumann measurement are the following
\begin{eqnarray}\label{12_36}P_{x+}&=&%\left(1/2\right)|1\rangle\langle1|=
I\otimes1/2\left(
                                             \begin{array}{cc}
                                               1 & 1 \\
                                               1 & 1 \\
                                             \end{array}
                                           \right),\quad
                      P_{x-}=%\left(-1/2\right)|2\rangle\langle2|=
                      I\otimes1/2\left(
                                             \begin{array}{cc}
                                               1 & -1 \\
                                               -1 & 1 \\
                                             \end{array}
                                           \right).
\end{eqnarray}
where $Tr P=2$, $P^2=P$.
Similarly, if we consider the observable $\sigma_z$ its eigenprojections are
\begin{eqnarray}\label{12_37}P_{z+}&=&I\otimes\left(
                                             \begin{array}{cc}
                                               1 & 0 \\
                                               0 & 0 \\
                                             \end{array}
                                           \right),\quad
                      P_{z-}=I\otimes\left(
                                             \begin{array}{cc}
                                               0 & 0 \\
                                               0 & 1 \\
                                             \end{array}
                                           \right).
\end{eqnarray}
\\ The evolution of the system is controlled by the unitary operator with the matrix $W$.
%\begin{eqnarray}\label{12_7}W = \left(
%                       \begin{array}{cccc}
 %                        u_{11} & u_{12} & u_{13} & u_{14} \\
 %                        u_{21} & u_{22} & u_{23} & u_{24} \\
 %                        u_{31}& u_{32} & u_{33} & u_{34}\\
  %                       u_{41} & u_{42} & u_{43} & u_{44} \\
  %                     \end{array}
  %                   \right).
%\end{eqnarray}
The state of the composite system after the interaction is the following
\begin{eqnarray*}\rho_{S+M}(k+1)=W\rho_{S+M}(k)W^{\dagger}
\end{eqnarray*}
and the reduced density matrix of the system we are interested in is
\begin{eqnarray*}
\rho_S(k+1)&=&Tr_M \rho_{S+M}(k+1)=Tr_M  W\rho_{S}(k)\otimes\rho_{M}(k)W^{\dagger}.
\end{eqnarray*}
The states after the measurement $\sigma_x$ or $\sigma_z$ are
\begin{eqnarray}\rho_{\pm}(k+1)&=&\frac{P_{\pm}\rho_{S+M}(k+1)P_{\pm}}{Tr\left(\rho_{S+M}(k+1)P_{\pm}\right)},\label{12_38}\end{eqnarray}
where $P_{\pm}$ are given by \eqref{12_36} or \eqref{12_37}, respectively.
%and the eigenstates of the measurement are
%\begin{eqnarray}\label{12_5}\theta_{S}(\pm1)&=&\frac{Tr_{M}\left(P_{\pm}\rho_{S+M}P_{\pm}\right)}{Tr\left(Tr_{M}\left(\rho_{S+M}P_{\pm}\right)\right)}
%=\frac{\widetilde{\rho}_{S\pm}}{Tr\left(Tr_{M}\left(\rho_{S+M}P_{\pm}\right)\right)}.
%\end{eqnarray}
\begin{example}
The evolution matrix can be taken as
$ W=e^{-ih(a_y\sigma_y^S\otimes\sigma_y^M)}$ (see \cite{Ruppert}), where $a_y$ is the coupling parameter  and $h$  is the sampling time. For
the measurement $A_x=I\otimes\sigma_x$ and  $a_yh=\pi/2$, the probabilities of two different outcomes are
\begin{eqnarray*}P(+1)=(1+\theta_{S_2}\theta_{M_3}),\quad P(-1)=(1-\theta_{S_2}\theta_{M_3}).
\end{eqnarray*}
The post measurement states are the following
\begin{eqnarray}\label{12_1}\theta_{S}(\pm1)&=&\left(
                                     \begin{array}{c}
                                       \frac{\theta_{S_3}\theta_{M_2}\pm\theta_{S_1}\theta_{M_1}}{1\pm\theta_{S_2}\theta_{M_3}} \\
                                        \frac{\theta_{S_2}\pm\theta_{M_3}}{1\pm\theta_{S_2}\theta_{M_3}} \\
                                        \frac{\pm\theta_{S_3}\theta_{M_1}-\theta_{S_1}\theta_{M_1}}{1\pm\theta_{S_1}\theta_{M_2}} \\
                                     \end{array}
                                   \right).
\end{eqnarray}
Since the probability of the new state depends on both measurements $\theta_{S}$ and $\theta_{M}$ we can retrieve the information about the useful state using only the observable measurement.
\end{example}
\begin{example} Let us observe the special case when $\theta_M(k)=[0,0,1]^T$, coupled with the system $\theta_S(k)=[0,0,1]^T$. Their density matrices are
\begin{eqnarray*}\rho_M(k)&=&\frac{1}{2}\left(
                                             \begin{array}{cc}
                                               \theta_{M_3}+1 & 0 \\
                                               0 & 1-\theta_{M_3} \\
                                             \end{array}
                                           \right)\equiv\left(
                                             \begin{array}{cc}
                                               R_{11} & 0 \\
                                               0 &  R_{22} \\
                                             \end{array}
                                           \right),\\
                      \rho_S(k)&=&\frac{1}{2}\left(
                                             \begin{array}{cc}
                                               \theta_{S_3}+1 & 0 \\
                                               0 & 1-\theta_{S_3}\\
                                             \end{array}
                                           \right)\equiv\left(
                                             \begin{array}{cc}
                                               r_{11} & 0 \\
                                               0 &  r_{22} \\
                                             \end{array}
                                           \right).
\end{eqnarray*}
The composite system is represented by
\begin{eqnarray*}\rho_{S+M}(k) = \left(
                      \begin{array}{cccc}
                        p_{11} & 0  & 0 &0 \\
                        0  & p_{22} & 0  & 0  \\
                        0 & 0  & p_{33} & 0 \\
                       0  &0  &0  & p_{44} \\
                      \end{array}
                     \right),
\end{eqnarray*}
where $p_{11}=R_{11}r_{11}$, $p_{22}=R_{11}r_{22}$, $p_{33}=R_{22}r_{11}$, $p_{44}=R_{22}r_{22}$.
Let us select the unitary evolution matrix of the following view
\begin{eqnarray}\label{12_8}W_{\varphi} = \left(
                       \begin{array}{cccc}
                            \cos{\varphi} & 0 & 0 & \sin{\varphi} \\
                           0 & 1 &0 & 0\\
                           0& 0 & 1 & 0\\
                            -\sin{\varphi} & 0 &0 &    \cos{\varphi} \\
                          \end{array}
                        \right).
   \end{eqnarray}
Hence, after the evolution step
\begin{eqnarray*}\rho_{S+M}(k+1) = \left(
                      \begin{array}{cccc}
                        p_{44}\sin^2{\varphi}+p_{11}\cos^2{\varphi} & 0  & 0 &-(p_{11}-p_{44})\sin{2\varphi}/2 \\
                        0  & p_{22} & 0  & 0  \\
                        0 & 0  & p_{33} & 0 \\
                       -(p_{11}-p_{44})\sin{2\varphi}/2  &0  &0  & p_{11}\sin^2{\varphi}+p_{44}\cos^2{\varphi} \\
                      \end{array}
                     \right),
\end{eqnarray*}
 for the observable $\sigma_z$ we use \eqref{12_38} and write
\begin{eqnarray}\rho_{z_{+}}&=&\frac{1}{p_{44}\sin^2{\varphi}+p_{11}\cos^2{\varphi}+p_{33}}\left(
                      \begin{array}{cccc}
                        p_{44}\sin^2{\varphi}+p_{11}\cos^2{\varphi} & 0  & 0 &0 \\
                        0  & 0 & 0  & 0  \\
                        0 & 0  & p_{33} & 0 \\
                       0  &0  &0  &0 \\
                      \end{array}
                     \right),\\
                      \rho_{z_{-}}&=&\frac{1}{p_{22}+p_{11}\sin^2{\varphi}+p_{44}\cos^2{\varphi}}\left(
                      \begin{array}{cccc}
                        0 & 0  & 0 &0 \\
                        0  & p_{22} & 0  & 0  \\
                        0 & 0  & 0 & 0 \\
                       0  &0  &0  &p_{11}\sin^2{\varphi}+p_{44}\cos^2{\varphi} \\
                      \end{array}
                     \right).\end{eqnarray}
                     If $\varphi=0$ one can rewrite the latter measurement as
          \begin{eqnarray}\rho_{z_{+}}&=&\frac{1}{p_{11}+p_{33}}\left(
                      \begin{array}{cccc}
                        p_{11} & 0  & 0 &0 \\
                        0  & 0 & 0  & 0  \\
                        0 & 0  & p_{33} & 0 \\
                       0  &0  &0  &0 \\
                      \end{array}
                     \right),\quad \rho_{z_{-}}=\frac{1}{p_{22}+p_{44}}\left(
                      \begin{array}{cccc}
                        0 & 0  & 0 &0 \\
                        0  & p_{22} & 0  & 0  \\
                        0 & 0  & 0 & 0 \\
                       0  &0  &0  &p_{44}\\
                      \end{array}
                     \right).\end{eqnarray}
%                     $\theta_{S}(k+1)=-\sin^2{\varphi}\theta_{S}(k)+\cos^2{\varphi}\theta_{M}(k)$.
It is clear that if the initial state is decohered and the evolutionary matrix is chosen such that at each step the matrix $\rho_{S+M}(\cdot)$ is diagonal, then we get the distribution function and its evolution at each step of the measurement. We can describe it in classical manner and apply the results known from the filtration theory discussed in Sec.~\ref{sec:12_0}.
\end{example}
%\par Let us select the unitary evolution matrix of the following view
%\begin{eqnarray}\label{12_8}W_{\varphi} = \left(
%                       \begin{array}{cccc}
   %                          \cos{\varphi} & 0 & 0 & \sin{\varphi} \\
   %                         0 & 1 &0 & 0\\
   %                          0& 0 & 1 & 0\\
    %                         -\sin{\varphi} & 0 &0 &    \cos{\varphi} \\
     %                      \end{array}
     %                    \right)
%    \end{eqnarray}
%The numerator and the denominator elements of \eqref{12_5} are given in appendix.
%    For simplicity, the angle $\varphi$ can be selected equal to zero and then the elements of \eqref{12_5} are
%    \begin{eqnarray*}
%    \widetilde{\rho_{S}}(1,1)&=&(\theta_{M_1} + 1)(\theta_{S_3} + 1)/4,\quad
%    \widetilde{\rho_{S}}(1,2)=(\theta_{S_1} - \theta_{S_2}i)(\theta_{M_1} + 1)/4,\\\nonumber
%    \widetilde{\rho_{S}}(2,1)&=&(\theta_{S_1} + \theta_{S_2}i)(\theta_{M_1} + 1)/4,\quad
%    \widetilde{\rho_{S}}(2,2)=-(\theta_{M_1} + 1)(\theta_{S_3} - 1)/4
%     \end{eqnarray*}
 %    and the denominator is $(\theta_{m_1}+ 1)/2$.
\subsection{Filtering of unknown quantum signal}
In the following, we assume that the Bloch's vector of unobservable qubit $\theta_{S_2}(k)$ is $s_k$, where $k$ is the time step.
In \cite{Ruppert} the observable ancilla qubit was characterized by the constant parameter  $c=\theta_{M_3}(k)$ and, repeating the given example,  the second row in \eqref{12_1} can be written as
\begin{eqnarray*}s_{k}&=&\frac{s_{k-1}\pm c}{1\pm c s_{k-1}}.
\end{eqnarray*}
Let us rewrite the latter process under the assumption that $c$ is small enough. Then the process will be the following
\begin{eqnarray*}s_{k}&=&s_{k-1}\pm c(1-s^2_{k-1})+O(c)
\end{eqnarray*}
or if we are interested in the system change only after $N$ time steps we get
\begin{eqnarray}\label{12_2}s_{k}&=&s_{k-1}+ x_{k-1}c(1-s^2_{k-1}),
\end{eqnarray}
where $x_k=x_{k+}-x_{k-}$, $N=x_{k+}+x_{k-}$ hold. The plus and the minus outcomes are denoted as $x_{k+}$ and $x_{k-}$, respectively.
\begin{rem}We can get the whole class of processes like \eqref{12_2} depending on the choice of the evolution matrix $W$.\end{rem}
In \cite{Ruppert} the latter equation was rewritten under the assumption that $c$ is small  and
 \begin{eqnarray*}x_k\sim \mathcal{N}\left(Ncs_k;N)\right)
\end{eqnarray*}
holds. This means that we can rewrite the measurement process as
 \begin{eqnarray}s_{k}&=&s_{k-1}+ Nc^2s_{k-1}(1-s^2_{k-1})+\omega_{k-1}c(1-s^2_{k-1}),\\\nonumber
 x_k &=& Ncs_k + \omega_k,\label{12_345}\end{eqnarray}
where the measurement noise is $\omega\sim\mathcal{N}\left(0;N)\right)$.
\par It is easy to see that we have the partially observable Markov random pair $(s_k,x_k)_{k\geq1}$, where the role of the unobservable sequence $s = (s_k)_{k\geq1}$  is played by the Bloch's vector of unobservable qubit $\theta_{S_2}$ and the observable sequence $x=(x_k)_{k\geq1}$ is obtained from von Neumann measurements. The connection between these variables is given by the nonlinear equation \eqref{12_2}.
Thus, a quantum observation model \eqref{12_345}, that is mathematically absolutely identical to \eqref{12_33}, is constructed. Further, it is possible to solve the problem of filtering for a given observation model  without thinking about the quantum nature of the observed and unobserved random sequences.
\begin{St}If the quantum systems is characterized by the nonlinear state equation \eqref{12_33}, where the observable sequence $x_k$ is obtained by means of the measurements performed on the ancilla qubit and $s_k$ is the unobservable sequence of measurements on the unknown system $S$, the optimal Bayesian estimator of the unknown $\widehat{s}_k$ that  minimizes the risk function \eqref{12_01} is provided by the optimal filtering equation \eqref{D}. \end{St}
\begin{rem} In \cite{Ruppert} the distribution of $x_k$ is approximated by the Gaussian distribution $x_k \sim \mathcal{N}(Ncs_k;N)$ and the Kalman-like filter is applied.
However, in general this approach does not provide  the optimal solution to the filtration problem.
We have to  use the optimal filtering equation \eqref{D} or its analogs for the non-exponential families of the densities since it gives the optimal solution of the filtration problem in both linear and non-linear cases as for linear as for non-linear problems.\end{rem}
\begin{example}
Let the quantum systems be characterized by the nonlinear state equation of the following form (see \cite{Dobrovidov:1983})
\begin{eqnarray*}b(x_{k})&=&s_{k}\eta_{k},
\end{eqnarray*}
where $b(\cdot)$ is a known differentiable function and the measurement noise is $\eta_{k}$.
Then our process looks like
 \begin{eqnarray}s_{k}&=&s_{k-1}+ b^{-1}(s_{k-1}\eta_{k-1})c(1-s^2_{k-1}),\\\nonumber
 x_{k}&=&b^{-1}(s_{k}\eta_{k}).\label{12_34}\end{eqnarray}
As a mathematical example, we assign  the observable $\eta_{k}$ distribution as
\begin{eqnarray*}p_{\eta}(y_k)=y_k^{t/2-1}e^{-y_k/2}/(2^{t/2}\Gamma(t/2))\quad \mbox{if} \quad y_k\geq0,
\end{eqnarray*}
where $t$ is a degree of freedom (a known number).
Then the conditional density $f(x_k|s_k)$
can be represented in the exponent form \eqref{exponetfamily}, where
\begin{eqnarray*}\widetilde{C}(s_k)=s_k^{-t/2+1}/(2^{t/2}\Gamma(t/2)),\quad h(x_k) = b(x_k)^{t/2-1},\quad T(x_k)= b(x_k)/2,\quad Q(s_k)=s_k^{-1},\end{eqnarray*}
where $Q(\cdot)$ is an invertible function and the optimal filtering equation \eqref{D} can be used to find the optimal Bayesian estimator $\mathsf{E}(Q(s_k)|x_1^k)$.
\end{example}
\section{Single - qudit observation model}\label{sec:12_2}
\par The above scheme of measurements is constructed on the basis of two qubits, i.e. the system with subsystems.
This is due to the fact that the quantum correlations and the entanglement phenomenon in the composite quantum systems
are viewed as a promising resource for the quantum technologies. However, the development of a universal toolbox for efficient control for large quantum systems
scalable with respect to number of subsystems is still a challenging problem of quantum science.
Recently, possibilities of using noncomposite quantum systems as a potential resource for the quantum information theory and in realizations of quantum technologies
have been discussed (cf. \cite{ManMan,Kessel}). The experimental demonstration of non-classical properties of noncomposite systems on the basis of a photonic qutrit is obtained in \cite{Lapkiewicz}. For the single - qudit system with the spin $j = 3/2$ the information and entropic
characteristics have been analyzed in \cite{Markovich3}.
\par  Thereby the  information properties of quantum states and their
characteristics could be associated not only with the composite quantum systems but also with the noncomposite ones.
Thus, all the quantum characteristics known for systems with subsystems can be mapped on the case of the systems without subsystems \cite{ManMan,Kessel,Markovich3,Mar8,Mar9}.
Then the single - qudit systems can be applicable in quantum technologies and therefore there is a need to construct the measurement  model for such kind of systems.
\par Motivated by this problem we try to extend the weak measurement scheme based on two qubits to the system of one single - qudit.
 The model can be formulated as follows. A single - qudit is the system of interest. The measurements are still produced  on the ancilla prepared qubit.
 However, such a model is similar to the two-qubit one and it is obvious that the Stratonovich's optimal nonlinear filtering method is applicable.
 We will consider another case.  Our experiment is built in such a way that there is a multi-level system in which some of the levels are accessible to measurements, and some are not. The question is, can we  "filter"  the information somehow at the inaccessible levels by the accessible ones?
\par Let the quantum state in four dimensional Hilber space $\mathcal{H}$ be described by the density matrix
\begin{eqnarray}\rho={\left(
                                 \begin{array}{cccc}
                                   \rho_{11}& \rho_{12}& \rho_{13}& \rho_{14}\\
                                   \rho_{21}& \rho_{22}& \rho_{23}& \rho_{24}\\
                                   \rho_{31}& \rho_{32}& \rho_{33}& \rho_{34}\\
                                   \rho_{41}& \rho_{42}& \rho_{43}& \rho_{44}\\
                                 \end{array}
                               \right)}\,. \label{33_1}
                               \end{eqnarray}
such that  $\rho=\rho^{\dagger}$, $Tr\rho=1$  and its eigenvalues are nonnegative.
The density matrix $\rho$ can describe the single - qudit system.  To this end we use the invertible mapping method introduced in \cite{Chernega:2008,Chernega:14,ManMan}.
 It was observed that quantum properties of the systems without
subsystems can be formulated using the invertible map of integers $1,2,3\ldots$ onto the pairs (triples, etc) of integers $(i,k)$, $i,k=1,2,\ldots$ (or semiintegers).
Thus, we rewrite $\rho$  using the following invertible mapping of indexes  $1\leftrightarrow 3/2$, $2\leftrightarrow1/2$, $3\leftrightarrow-1/2$, $4\leftrightarrow-3/2$ as
\begin{eqnarray*}\rho_{3/2}&=&\left(
                                 \begin{array}{cccc}
                                   \rho_{3/2,3/2}& \rho_{3/2,1/2}& \rho_{3/2,-1/2}& \rho_{3/2,-3/2}\\
                                   \rho_{1/2,3/2}& \rho_{1/2,1/2}& \rho_{1/2,-1/2}& \rho_{1/2,-3/2}\\
                                   \rho_{-1/2,3/2}& \rho_{-1/2,1/2}& \rho_{-1/2,-1/2}& \rho_{-1/2,-3/2}\\
                                   \rho_{-3/2,3/2}& \rho_{-3/2,1/2}& \rho_{-3/2,-1/2}& \rho_{-3/2,-3/2}\\
                                 \end{array}
                               \right).
                               \end{eqnarray*}
 The matrix saves the standard properties of the density matrix, i.e. $\rho_{3/2}=\rho_{3/2}^{\dagger}$, $Tr\rho_{3/2}=1$ hold and its eigenvalues are nonnegative.
Using the partial trace two "artificial subsystems" (the "artificial qubits") can be introduced as
\begin{eqnarray}\rho_1={\left(
                            \begin{array}{cc}
                              \rho_{3/2,3/2}+ \rho_{1/2,1/2}& \rho_{3/2,-1/2}+ \rho_{1/2,-3/2}\\
                               \rho_{-1/2,3/2}+\rho_{-3/2,1/2}&\rho_{-1/2,-1/2}+\rho_{-3/2,-3/2}\\
                            \end{array}
                          \right)}\, , \label{33_7}
\end{eqnarray}
\begin{eqnarray}
                          \rho_2={\left(
                            \begin{array}{cc}
                              \rho_{3/2,3/2}+\rho_{-1/2,-1/2} & \rho_{3/2,1/2}+ \rho_{-1/2,-3/2}\\
                              \rho_{1/2,3/2}+\rho_{-3/2,-1/2} & \rho_{1/2,1/2}+\rho_{-3/2,-3/2} \\
                            \end{array}
                          \right)}\,. \label{33_77}
\end{eqnarray}
Then we think that the first "artificial subsystems" corresponds to the unobserved qubit of interest and the second "artificial subsystems" corresponds to the ancilla qubit.
\par Hence, one can think that the density matrix  $\rho_{S+M}(k)$ describes the single - qudit state. Applying the latter method, the two  "artificial subsystems" $\rho_{S}(k)$ and $\rho_{M}(k)$ can be constructed such that $\rho_{S+M}(k)=\rho_{S}(k)\otimes\rho_{M}(k)$. Then one can construct the observation model as
 for the two-qubit system. However, it does not contain the real observable ancilla qubit  and the unobservable one.
The described method gives the possibility to extend many known quantum characteristics such as entanglement and steering to systems without subsystems (see \cite{Mar8,Mar9}).
Note that we started our reasoning with the density matrix of a general form. Thus, a similar construction holds for any qudit density matrix. It is possible to construct other partitions into artificial subsystems and thus construct other measurement models for a system of one qudit. The difference is that in this case the role of the ancilla system is played by the accessible energy levels of the multilevel atom and the system of interest are unaccessible levels. The filtering method provides an elegant solution of getting information about  unaccessible  levels by accessible levels. Thus, applying the filtration method  to the single - qudit system, we solve a completely different problem with another physical meaning and the observable random variables.

\subsection{Physical meaning of the "artificial qubits"}
Let us start from the short example of two coins which can drop on the first ($1$) or on the second ($2$) side.
Hence, there are two random variables $m_1$, $m_2$  and four events $(m_1,m_2)=\{(11), (12), (21), (22)\}$ with probabilities $p_{ij},\quad i,j=1,2$.
Let $\omega(m_1,m_2)$ be the joint probability function of these two random variables. Their marginal probability functions can be defined as
 \begin{eqnarray*}\omega_1(m_1)&=&\sum\limits_{m_2}\omega(m_1,m_2),\quad
  \omega_2(m_2)=\sum\limits_{m_1}\omega(m_1,m_2).\end{eqnarray*}
 Hence, the correlation between the two observations is given by
 \begin{eqnarray*}\langle m_1,m_2\rangle=\sum\limits_{m_1,m_2}m_1m_2\omega(m_1,m_2).\end{eqnarray*}
 \par However, we can be interested not in the whole system, but only in  cases when the first coin falls on the first side and the second coin is not interesting for us.
 We have two new events $\{\widetilde{\omega}\}$ with probabilities  $\widetilde{p}_1=p_{11}+p_{12}$ and $\widetilde{p}_2=p_{22}+p_{21}$.
 Analogically,  if we are interested only in the second coin side, we have two new events $\omega$ with probabilities $p_1=p_{11}+p_{21}$ and $p_2=p_{22}+p_{12}$.
 Outcomes $\{\omega\}$ and $\{\widetilde{\omega}\}$
  are correlated. The latter example  shows the existence of correlations in the systems without subsystems.
  \par If we have a single - qudit system with the spin $j=3/2$, we can write the sample space $\Omega$ with four events $\omega\in\Omega$ and values of  spin projections
  $|m\rangle=\{|3/2\rangle, |1/2\rangle, |-1/2\rangle, |-3/2\rangle\}$ with probabilities $p_{3/2},p_{1/2},p_{-1/2},p_{-3/2}$. Then we have one four-level atom, e.g.,
$|m\rangle=|3/2\rangle$  corresponds to the case when the highest (fourth) level is filled.
If we are interested only in  outcomes when the fourth or the second levels of the four-level atom are filled, then we can assume that we have a new set of two
 outcomes $\{\omega_1\}$ with  probabilities $p_1=p_{3/2}+p_{-1/2}$,  $p_2=p_{-3/2}+p_{1/2}$.
 If we are interested only in outcomes when the fourth and the third levels are filled, then we have another set of
  outcomes $\{\omega_2\}$ and their probabilities are $\widetilde{p}_1=p_{3/2}+p_{1/2}$,  $\widetilde{p}_2=p_{-3/2}+p_{-1/2}$.
  The outcomes $\{\omega_1\}$ and $\{\omega_2\}$ are correlated. Hence,  correlations in single - qudit systems are between different combinations of outcomes.
  \begin{rem} With respect to our problem, the experiment could be designed so that we can measure the population only on certain  levels, while others are not available to measure.
  This may be due to their short lifetimes or a diversity in the frequency band reception.
  For example, we can measure only the first and the second levels of the four-level atom and the third and the fourth levels are unobservable.
  Then we can think about the observable levels as about "artificial ancilla qubit" and about other two levels as about "artificial unobservable qubit".
  Thus, we can construct the observation model just like for the real two-qubit system.\end{rem}
\section{Conclusion}\label{sec:12_5}
\par To conclude let us point out the main results of our work. Using  the known state space model of a particular quantum
system based on the weak measurements we demonstrated how to apply classical filtration methods in this case.
Since  the Kalman filter approach does not give the optimal solution for nonlinear models, we
propose  to use for nonlinear quantum models the general filtering equation. The latter gives the optimal Bayesian solution.  In contrast to the known state estimation methods we do not need any simplifications or linearizations of the state space model.
\par Moreover, using the invertible map of indices we extend the  observation model based on the indirect measurement
 known for the two-qubit system to the single - qudit system. This observation  model is useful in case when it is experimentally impossible to do measurements at all levels of the multilevel system. Then we can "filter" the information at inaccessible levels by conducting measurements on observable ones. Therefore, having the nonlinear observation model one  can forget about its physical nature and apply the filtering method described above.
% Finally, using the density matrices of the unknown and the ancilla states, we introduce the tomograms of these states and their joint state and show how the tomogram changes over  time steps. The tomographic approach is also used to write the Shannon entropies for the examined states depending on the time step of the quantum observation model.
%Using these entropies, the information inequality  dependent on time step is obtained. The time evolution of the quantum information illustrates the  correlation changes in the quantum system.
\par The main advantage of the proposed nonlinear filtration method in both classical and quantum models is its optimality. The general filtering  equation  gives better result in comparison with any linearization methods and all Kalman-like filters because of the proven optimality in the sense of minimal Bayesian risk. The proposed method has drawbacks, such as the dependence on the unknown conditional density $f(x_k|x_1^k)$. This can be estimated, for instance, using the nonparametric kernel density estimation methods (see \cite{Mar_1,Mar_2}). A possible development of the introduced ideas can be a consideration of qudits  and the construction of a nonlinear observation model for such a system with the further application of a nonlinear filter.
\section*{Acknowledgements}
L.A.M.  was partly supported by the Russian Foundation for Basic Research, grant 13-08-00744 A

\end{document}